\documentclass[11pt]{article}
\usepackage{fa2025}
\usepackage{amsmath}
\usepackage{cite}
\usepackage{url}
\usepackage{graphicx}
\usepackage{color}
\usepackage{siunitx}
\usepackage[utf8]{inputenc}
\usepackage{soul} 
\usepackage{amsmath}
\usepackage{amssymb}
\newcommand{\R}{\mathbb{R}}

\newcommand{\vect}[1]{\boldsymbol{#1}}

\newcommand{\xmark}{\color{red} \ding{55}}
\newcommand{\cmark}{\color{green} \ding{51}}

\definecolor{topcolor}{RGB}{161, 124, 169} 
\definecolor{middlecolor}{RGB}{105, 143, 166}  
\definecolor{bottomcolor}{RGB}{105, 143, 166}

\newcommand{\mvc}{\color{middlecolor}} 
\newcommand{\tvc}{\color{topcolor} }
\newcommand{\bvc}{\color{bottomcolor}}

\usepackage{booktabs}   
\usepackage{array}      
\usepackage{graphicx}   
\usepackage{caption}    
\usepackage{siunitx}  

\usepackage{pifont}

\title{Towards Interpretable Emotion Recognition: Identifying Key Features with Machine Learning}

\multauthor
{Yacouba Kaloga$^{*}$ \hspace{1cm} Ina Kodrasi 
} {
\\
  Signal Processing for Communication Group, Idiap Research Institute, Martigny, Switzerland\\
\correspondingauthor{yacouba.kaloga@idiap.ch}{Kaloga et al.}
}
\begin{document}
\maketitle
\begin{abstract}
Unsupervised methods, such as wav2vec2 and HuBERT, have achieved state-of-the-art performance in audio tasks, leading to a shift away from research on interpretable features. However, the lack of interpretability in these methods limits their applicability in critical domains like medicine, where understanding feature relevance is crucial. To better understand the features of unsupervised models, it remains critical to identify the interpretable features relevant to a given task. 
In this work, we focus on emotion recognition and use machine learning algorithms to identify and generalize the most important interpretable features for this task. While previous studies have explored feature relevance in emotion recognition, they are often constrained by narrow contexts and present inconsistent findings. Our approach aims to overcome these limitations, providing a broader and more robust framework for identifying the most important interpretable features.
\end{abstract}

\section{Introduction}\label{sec:introduction}

Effective communication relies on the accurate mutual perception of emotions between interlocutors. A significant portion of 
 information exchanged in a conversation is conveyed through emotional cues—visually, as well as through voice, intonation, and other auditory features. When emotional perception is impaired, as in certain pathologies~\cite{leung2023emotion}, communication becomes particularly challenging, underscoring the importance of preserving these cues in auditory interactions.
At the same time, a growing share of our audio interactions now occurs through electronic systems, including remote meetings, phone calls, and hearing aids. While these systems process sound to optimize factors such as signal quality, signal intelligibility, energy efficiency, or latency, the preservation of emotional cues is rarely considered.  Evaluating the impact of various processing schemes on emotional cues through human subjective testing is impractical, due to high costs and limited scalability. Automated systems that use machine learning models for speech emotion recognition (SER) across different sound processing schemes can provide a viable alternative. Previous studies have shown similarities between human and machine emotion perception, with machines typically relying on handcrafted acoustic features to identify key emotional cues~\cite{lim14,cou12}. The impact of various processing schemes (or perturbations) on emotion perception can then be quantified either by analyzing how the handcrafted acoustic features are affected~\cite{par23,koe20} or by measuring performance degradation in models that depend on these features~\cite{par23,par18}.
Such approaches, however, typically rely on a narrow range of acoustic features and limited number of datasets, lacking generalizability and highlighting the need for a broader set of emotionally relevant acoustic features that can be applied across various settings.

Our work aims to identify relevant acoustic features for SER using various automatic classifiers on various datasets. More specifically, we employ six classification models, a significantly higher number compared to what is typically seen in the SER literature.
Additionally, we use six different datasets in four different languages, with models trained multiple times using different splits of the same dataset.
Utilizing multiple models (trained multiple times on different splits of the data) and datasets provides several advantages. First, extracting the relevant acoustic features across multiple models (resp. datasets) increases robustness. If a particular feature is consistently important across different models (resp. datasets), it indicates its reliability and reduces the influence of individual model (resp. dataset) characteristics. Additionally, using multiple models helps mitigate the influence of underperforming models whose feature importances are unreliable due to poor performance, as well as models that overperform by relying on features highly specific to a particular dataset. Secondly, testing feature importance algorithms on multiple models and different datasets enables us to gauge the generalizability of feature rankings across different learning paradigms and datasets. Finally, running the analysis multiple times can help mitigate the challenges posed by correlated features.
Specifically, our contributions are as follows:

\begin{itemize}
    \item  We derive the most relevant acoustic features for SER across different classification models and datasets.
    \item  We propose an approach to combine the relevance of features from each individual model and dataset, resulting in a single, robust, and generalizable list of key acoustic features.
    \item We experimentally demonstrate the advantages of our approach for establishing key acoustic features, showing that it outperforms using a single classifier or a single dataset in terms of robustness and generalizability.
\end{itemize}

\section{Related work}\label{sec:speech_emo_rec}

\textit{Research on speech emotion perception has primarily explored key acoustic features and their impact on human perception. In this section, we review these studies and their limitations. Further, we examine how machine learning can offer more general findings by comparing its performance to human perception. Finally, we discuss why current studies leveraging this connection remain limited.}

\subsection{Acoustic Cues and Human Emotion Perception}\label{sec:acoustic}

Numerous studies have established a strong connection between specific acoustic features and human perception of emotions. Pitch variations play a crucial role, with studies showing that higher mean pitch values are associated with high-arousal emotions such as anger, fear, and excitement, while lower values are linked to low-arousal emotions like boredom~\cite{moz98, sch06, luo07, ham07, rod11}.
Temporal aspects, such as speech rate and rhythm, and spectral measures, including formants and spectral flux, are also critical in conveying emotions~\cite{lim14, moz01}. Additional acoustic cues, such as prosody contour, intensity, and energy-related features, further enrich the emotional expressiveness of speech~\cite{guz13, wu09}. Importantly, these acoustic features do not function in isolation but interact in complex ways to shape emotional perception.
Despite these insights, existing studies face limitations. Most experiments rely on small sample sizes due to the high cost and time required for participant-based evaluations. This leads to findings that are highly context-dependent and difficult to generalize~\cite{luo07}. Additionally, some studies report contradictory results, likely due to differences in methodology, dataset composition, and speaker variability (for example, ~\cite{sch18} has inconsistent findings on sadness with respect to \cite{par17,tho06,tsi10,pel09}).
Given these constraints, machine learning offers a promising avenue for studying emotional perception, as it allows for the analysis of large-scale datasets and the identification of underlying patterns that may be difficult to discern from human-limited studies.

\subsection{Machine-based SER and Human Emotion Perception}

Recent studies have directly compared machine learning models to human emotion perception. For instance, ~\cite{par23} evaluated classification models (i.e., multilayer perceptron (MLP) and support vector machine (SVM)) against human perception under both clean and noisy conditions. Their study introduced emotion distractors, i.e., emotions present in the test set but unseen during training, to ensure that models were performing true recognition rather than simple discrimination. Their results showed striking similarities between human and machine perception, including comparable overall SER performance, similar relative accuracy across different emotions, consistent performance degradation in noisy environments, and similar confusion matrices, particularly for well-defined emotions like anger and sadness. These findings align with earlier literature~\cite{par18, cou12}, which also reported strong similarities between human and machine SER. Additionally, regression models predicting valence/arousal positioning have shown that human-based emotion placement can effectively predict machine performance on the same task~\cite{lim14}. This suggests that both humans and machine learning models rely on similar acoustic cues when interpreting emotions~\cite{koe20, par23, cou12, jeon13}.

\subsection{Acoustic Cues and SER}\label{sec:acoustic}

To the best of our knowledge, a limited number of studies have quantitatively explored the most important acoustic cues for SER in machine learning-based models. Such work is carried out using mainly regression and fixed effect regression \cite{sch18,lim14,koe20}. Since features importance results may typically depend on the classifier/regressor use, the generalizability of their conclusions is limited. 
Additionally, the number of features considered is relatively low, and some key features may be overlooked in these analyses. The study context is often limited to a small number of speakers, a few datasets, and a restricted linguistic environment. In~\cite{jeon13}, it is highlighted that models perform worse than humans in cross-dataset settings, emphasizing the lack of generalizability in these constrained setups. Furthermore, due to individual authors’ choices—such as using different emotions or acoustic cues—aggregating findings to draw broader conclusions remains challenging.
Given these limitations, our objective is to derive the most important acoustic cues for SER that generalize well across multiple representative datasets and models.

\section{Proposed Approach}\label{sec:speech_emo_rec}

We consider $D$ datasets, each consisting of $n_d$ utterances $\vect{x}_i$, $i = 1, \ldots, n_d$.  Each utterance $\vect{x}_i$ is associated with an emotion label $y_i$, such as e.g., {\emph{happiness}}, {\emph{anger}}, {\emph{sadness}}, etc. 
The SER task is a supervised task, where we train a model to learn to predict the correct emotion label $y$ of an input utterance $\vect{x}$. 
In this section, we propose an approach for identifying the most  relevant acoustic features for SER using various datasets and various models.

\subsection{Feature Set}

Each utterance \(\vect{x}_i\) is represented by a fixed-size feature list \(f = [f_1(\vect{x}_i), \dots , f_{Q}(\vect{x}_i)] \in \R^{Q}\), designed to capture diverse acoustic characteristics.
To construct this feature list, we use the \textit{ComParE\_2016} feature set~\cite{eyb10}, a large collection of features (\(Q = 6373\)) not specifically designed for SER but known to capture a broad range of acoustic information. This set is based on extracting $65$ low-level descriptors (LLDs) such as e.g., pitch, spectral energy, and auditory spectrum, along with an additional $65$ LLDs derived as their temporal differences. 
Since LLDs are time-varying and utterances are of different lengths, a wide range of functional descriptors are then applied to these LLDs (such as various distribution moments, local extrema, and different types of quantiles) in order to construct the fixed-size feature list.

\subsection{Deriving Key Acoustic Features}
\label{sec:der}
To identify the most important acoustic features for SER, we use classification models that inherently provide feature importance scores, quantifying each feature's contribution to the models' decision.
To reliably leverage these importance scores to derive the most important acoustic features, achieving strong classification performance is essential. Additionally, special consideration must be given to correlated features. In many feature importance models, a single variable may be assigned high importance while its correlated counterparts, despite being equally relevant, may be overlooked. Conversely, importance may be spread across a cluster of correlated variables, creating the misleading impression that individual features lack significance. Addressing this challenge is a key aspect of feature importance analysis. tm This is also why our approach, which involves multiple models, datasets, and experiment repetitions, is relevant. If two features convey the same information, they should exhibit similar average or median importance across all these experimental conditions.

To identify the most important acoustic features for SER, we use $M$ classification models and $K$ datasets.
While $M=4$ and $K=6$ are used in Section~\ref{sec:fs}, the proposed approach is applicable to any number of models $M$ and number of datasets $K$.
Every time a model $m$ is trained on a dataset $d$, with $m = 1, \ldots, M$ and $d = 1, \ldots, D$, the outcome is the classification performance $p_{d}^{m}$ and the importance score of each feature $s_{q,d}^{m}$.
Preliminary experiments have shown that aggregating all feature importance scores in one step, such as e.g., using \( s_q = \frac{1}{D M} \sum s_{q,d}^{m} \), is inefficient. Specifically, the resulting selection of features is not as effective as using the selection of features from a single model with respect to the evaluation metrics introduced in Section~\ref{fie}.
Instead, we propose to find the key acoustic features for SER by proceeding as following.

First, we aggregate the normalized feature importance scores across all considered models for each individual dataset. The aggregated importance scores $s_{q,d}$ are computed as
\begin{equation}
s_{q,d}  = \text{median}\Bigg\{\frac{s_{q,d}^{m}}{\max\limits_{q=1}^{\text{Q}} \big\{s_{q,d}^{m}\big\}}\Bigg\}_{m=1}^{M}. 
\end{equation}
Then, for each dataset $d$, we construct a new feature list \( f^{\text{order}}_{d} \) by ordering the features in $f$ in decreasing order of importance according to the scores $s_{q,d}$.
Each model is then retrained using subsets of the top-ranked features, varying from $0.5$\% to $20$\% in increments of $0.5$\%.
This process is stopped as soon as the model achieves a performance that matches or exceeds the original performance \(\ p_{d}^m \) obtained using all features.
Let \( pt(m) \) denote the minimum percentage of top-ranked features required for model 
$m$ to reach at least par performance. We then define the multiset $M$ (allowing repeated elements) of all features that were used to reach this threshold
\begin{equation}
M = \{f_{q,d}^{\text{order}} \mid \forall q \in [\!|1,Q |\!], \forall m \in [\!|1,\text{M} |\!],  q \leq pt(m)\}.
\end{equation}
Since the features correspond to statistics of LLDs, we identify the most important acoustic cues for SER by counting the occurrences of each LLD statistic in the multiset 
$M$, and ranking them based on frequency.

\section{Experimental Settings}\label{sec:xp_set}

\textit{In this section, we present the SER datasets, classification models, and feature importance algorithms employed. Finally, we conclude this section with a detailed description of model training and evaluation.}

\subsection{Datasets and Emotions}

To derive a generalizable list of important acoustic cues, we employ six datasets commonly used in the speech emotion recognition (SER) literature, i.e., \emph{CaFE}\cite{gou18}, \emph{Emovo}\cite{cos14}, \emph{EmoDB}\cite{bur05}, \emph{DEMoS}\cite{emi19}, \emph{Tess}\cite{pic20}, and \emph{Gemep-5emo}\cite{ban12}.
As summarized in Table~\ref{tab:datasets}, these datasets span four languages—Italian, German, French, and English.
According to the emotion mapping shown in Table~\ref{tab:new_emotions}, we consider utterances labeled as \emph{neutral}, as well as those representing six core emotions, i.e., \textit{happiness}, \textit{(hot) anger}, \textit{fear}, \textit{sadness}, \textit{surprise}, and \textit{disgust}.

\begin{table*}[t!]
\caption{Summary of the used emotion recognition datasets and their characteristics. 
Neutral utterances denote utterances with an emotionally neutral meaning, chunks refer to segments extracted (based on syntax or prosody of the speaker) from sentences, whereas non-sense utterances are plausible sequences of language sounds without any meaning. The Male (F) column summarizes the number of male (female) speakers in each dataset. French C. denotes French Canadian. The Utterances column presents the total number of utterances in each dataset ($\#$), the number of distinct sentences uttered in various emotions when applicable ($\#$ Sentences), and the mean ($\mu$) and standard deviation ($\sigma$) of the duration of utterances (in seconds) in each dataset.}
\scalebox{0.72}{
\centering
    \centering
    \renewcommand{\arraystretch}{1.4} 
    \setlength{\tabcolsep}{6pt} 
    \begin{normalsize} 
    \begin{tabular}{@{}cccccccccc@{}}
        \toprule
         & \multicolumn{3}{c}{\textbf{Datasets Characteristics}} & \multicolumn{3}{c}{\textbf{Speakers}} & \multicolumn{3}{c@{}}{\textbf{Utterances}} \\
        \cmidrule(r){2-4} \cmidrule(lr){5-7} \cmidrule(l){8-10}
        \textbf{Datasets}& \textbf{Language} & \textbf{Acted} & \textbf{Utterance Types} & \textbf{$\#$ Male (F)} & \textbf{Native Tongue} & \textbf{Actor} & \textbf{$\#$} & \textbf{$\#$ Sentences} & \textbf{Duration $\mu$/$\sigma$ (s)} \\
        \midrule
        \textbf{\textbf{CaFE}} & French & \cmark & Neutral & 6 (6) & French C. & \cmark & 936 & 6 & 4.4/0.8 \\
        \textbf{DEMoS} & Italian & Induced & Chunks & 45 (23) & Italian & \xmark 
        & 9697 & - & 2.9/1.3 \\
        \textbf{\textbf{EmoDB}} & German & \cmark & Neutral & 5 (5) & German & \cmark & 494 & 10 & 2.8/1.0 \\
        \textbf{\textbf{Emovo}} & Italian & \cmark & Neutral & 3 (3) & Italian & \cmark & 588 & 14 & 3.1/1.4 \\
        \textbf{\textbf{Gemep-5emo}} & French & \cmark & Non-sense & 5 (5) & French & \cmark & 50 & 2 & 2.6/1.2 \\
        \textbf{\textbf{Tess}} & English & \cmark & Neutral word & 0 (2) & English & \cmark & 2800 & 200 & 2.1/0.3 \\
        \bottomrule%
    \end{tabular}
    \end{normalsize}
}

\label{tab:datasets}
\end{table*}

\begin{table}[t!]
\caption{Summary of the emotions and number of utterances for each emotion in the considered datasets.
\label{tab:new_emotions}}
\hspace{-1.2cm}
\scalebox{0.7}{
\centering
    \begin{tabular}{@{}ccccccccccc@{}}
        \toprule
        & \multicolumn{7}{c}{\textbf{Emotion}} \\
        \cmidrule(r){2-8}
        \textbf{Datasets} & \textbf{Happiness} & \textbf{Anger} & \textbf{Fear} & \textbf{Sadness} & \textbf{Neutral} & \textbf{Disgust} & \textbf{Surprise} \\
        \midrule
        \textbf{\textbf{CaFE}} & 120 & 120 & 120 & 120 & 60 & 120 & 120  \\
        \textbf{DEMoS} & 1395 & 1477 & 1156 & 1530 & 332 & 1678 & 1000 \\
        \textbf{\textbf{EmoDB}} & 71 & 127 & 69 & 62 & 79 & 46 & \xmark  \\
        \textbf{\textbf{Emovo}} & 84 & 84 & 84 & 84 & 84 & 84 & 84 \\
        \textbf{\textbf{Gemep-5emo}} & 10 & 10 & 10 & 10 & \xmark & \xmark & \xmark  \\
        \textbf{\textbf{Tess}} & 400 & 400 & 400 & 400 & 400 & 400 & \xmark \\
        \midrule
        {\bf{Total}} & 2080 & 2218 & 1839 & 2206 & 955 & 2382 & 1204 \\ 
        \bottomrule
    \end{tabular}

}

\end{table}

\subsection{Performance Evaluation}

The standard evaluation metric used in the SER literature is the Unweighted Average Recall (UAR), which is derived from the individual recalls of each emotion. A model's recall for a specific emotion is calculated by dividing the number of correctly identified instances of that emotion by the total number of instances of that emotion in the dataset. This metric is preferred over the conventional accuracy metric since it focuses on the model's capability to recognize a particular emotion without emphasizing discrimination among other emotions. The UAR is simply the arithmetic mean of the recalls across all emotions\footnote{Also known as the Macro Recall.}.

\subsection{Models}

The classifiers used to compute the most important acoustic features for SER include the Linear Support Vector Machine (L-SVM) \cite{vapnik}, Logistic Regression (LGRG) \cite{hosmer}, Random Forest Fourier (RFC) \cite{NIPS2007}, and Light Gradient Boosting Machine (LGBM) \cite{Ke_2017} classifiers. These classifiers were chosen for their inherent ability to determine feature importance. Findings are then validated using two additional classifiers, i.e., the Radial Basis Function SVM (RBF-SVM) \cite{vapnik} and Multilayer Peceptron (MLP) \cite{rumelhart1986learning} (which do not offer a straightforward method for determining feature importance). These classifiers were selected because they exhibit similarities to human perception in SER~\cite{par23}. Each classification model has a set of hyperparameters, which we optimize using grid search focused on key parameters, balancing training efficiency with strong performance (cf. Section~\ref{sec: hyper}).

\subsection{Selection of Hyperparameters}
\label{sec: hyper}
In this section, we outline the process of selecting optimal hyperparameters for each considered model. The selection is carried out in three key steps, i.e.,:

\begin{enumerate}
  \item  \emph{Split datasets.} \enspace To ensure an unbiased evaluation with respect to speaker and sex specificity, we carefully split each dataset into training and test sets based on two key principles. First, all utterances from a given speaker are assigned exclusively to either the training or the test set, preventing any speaker from appearing in both. Second, we strive for the best possible male-female balance among speakers in the test set.
Following these principles, the splitting strategy is as follows: for datasets with more than $12$ speakers, 20\% of the speakers are assigned to the test set. For datasets with fewer than $12$ speakers, two speakers are assigned to the test set. Exceptions are made for \emph{Emovo} (with six speakers in total) and \emph{Tess} (with two speakers in total), where only one speaker is used for testing.

 \item \emph{Validation.} \enspace 
 To select the optimal hyperparameters, we use stratified $5$-fold cross-validation. For each hyperparameter configuration $p$, the model is trained on four folds and evaluated on the remaining fold. This process is repeated five times, ensuring that each fold serves as the validation set once. The results from these five runs are then averaged, and the configuration yielding the highest average performance is selected as the optimal hyperparameter configuration $p^{*}$.
 This cross-validation strategy enables robust assessment of different hyperparameter combinations and ensures effective selection for each model.
 For the larger DEMoS dataset, due to its size, a single evaluation is performed using one fold as a dedicated validation set.

\item \emph{Test.} \enspace Finally, we train each model on the entire training set using the hyperparameters $p^{*}$ that were determined in step $2$. Then, we evaluate this trained model on the test set to assess its performance on unseen data.
\end{enumerate}
To further reduce bias, this procedure is repeated $3$ times for each dataset (i.e., using $3$ different splits of the data into training and test sets), and the average of these three runs corresponds to the model's performance on that particular dataset.

\begin{table*}[h]
\caption{Mean and standard deviation of the UAR across three runs for each considered model and dataset. We highlight in purple all values within one percent of the best value in each row.}
\centering
\scalebox{0.9}{
\hspace{-0.9cm}
\begin{tabular}{@{}c|cccccc|c@{}}
\toprule
\textbf{Datasets} & \textbf{RBF-SVM} & \textbf{MLP} & \textbf{RFC} & \textbf{LGBM} & \textbf{L-SVM} & \textbf{LGRG} & \textbf{Avg. across models} \\
\midrule
\textbf{CaFE}   & \mvc 48.6$\pm$4.0 & \tvc 50.6$\pm$2.8 & \bvc 40.1$\pm$1.6 & \mvc 44.6$\pm$4.0 & \tvc 49.8$\pm$3.8 & \tvc 49.6$\pm$2.6 & \mvc 47.2$\pm$3.7 \\
\textbf{EmoDB}  & \mvc 78.8$\pm$2.4 & \mvc 80.3$\pm$1.5 & \mvc 70.8$\pm$0.9 & \mvc 75.3$\pm$0.1 & \tvc 82.7$\pm$2.3 & \tvc 83.9$\pm$1.8 & \mvc 78.6$\pm$4.5 \\
\textbf{DEMoS}  & \tvc 74.2$\pm$0.1 & \tvc 74.9$\pm$0.8 & \mvc 51.6$\pm$0.6 & \mvc 63.8$\pm$0.8 & \mvc 65.9$\pm$0.3 & \mvc 66.0$\pm$0.3 & \tvc 66.1$\pm$7.7\\
\textbf{Emovo}  & \tvc 43.2$\pm$5.8 & \bvc 36.7$\pm$5.0 & \tvc 43.5$\pm$1.7 & \tvc 44.2$\pm$3.3 & \mvc 36.1$\pm$5.1 & \bvc 35.7$\pm$4.7 & \mvc 39.9$\pm$3.8 \\
\textbf{Gemep-5emo} & \tvc 83.3$\pm$8.2 & \mvc 80.0$\pm$14.1 & \mvc 80.0$\pm$12.2 & \mvc 76.7$\pm$8.2 & \mvc 80.0$\pm$7.1 & \mvc 76.7$\pm$4.1 & \mvc 79.5$\pm$2.3 \\
\textbf{Tess}   & \mvc 58.9$\pm$0.0 & \mvc 61.7$\pm$0.9 & \mvc 57.0$\pm$0.4 & \mvc 57.6$\pm$0.4 & \tvc 69.9$\pm$0.0 & \tvc 69.6$\pm$0.0 & \mvc 62.5$\pm$5.4 \\
\midrule
\textbf{Average across datasets} & 
\tvc 64.5$\pm$15.2 & 
\tvc 64.0$\pm$16.2 & 
\mvc 57.2$\pm$14.2 & 
\mvc 60.4$\pm$13.0 & 
\tvc 64.1$\pm$16.5 & 
\tvc 63.6$\pm$16.3 & 
62.5$\pm$15.5 \\
\bottomrule
\end{tabular}
}
\label{tab:experiments}
\end{table*}

\subsection{Features Importance Evaluation}
\label{fie}
In order to validate the procedure above in the experimental part, we need to be able to determine which feature importance list is better. To do so, we rank the features in order of importance for both lists. We then retrain our model using a growing percentage of features in their importance order. The feature selection that achieves the same performance as using all features with a smaller percentage of features will be considered better.

\section{Experimental Results}\label{sec:xp_res}

\textit{In this section, we present the results of the experiments described previously, followed by an analysis of the most important features. Note that when referring to the selection of important features, we mean selecting those with the highest importance scores from a ranked list of feature importance coefficients.}

\subsection{Performance of all models across for all datasets using the complete feature set}
\label{model_performance}

Table~\ref{tab:experiments} summarizes the unweighted average recall (UAR) achieved by each model on the considered datasets, using the training and validation procedure described in Section~\ref{sec: hyper}.

From a dataset-level perspective, it is evident that classification performance is significantly influenced by the specific dataset. Notably, the number of speakers appears to affect the variance observed across the three runs (each involving different speaker combinations in the test set). Datasets with fewer speakers generally show higher variance (with the Tess dataset being a notable exception). This increased variability may stem from the limited number of speakers, which restricts the model’s ability to generalize and leads to overfitting on speaker-specific traits present in the training data. This observation highlights the importance of using a diverse and extensive set of datasets, as done in this study.
Beyond dataset size, no clear correlations are observed between classification performance and other dataset characteristics (as listed in Table~\ref{tab:datasets}), such as the speakers’ native language, the spoken language, or the nature of the utterances.

From a model-level perspective, the RBF-SVM, L-SVM, MLP, and LGRG models generally yield the best overall performance. However, on the EMOVO dataset, the RFC and LGBM models considerably outperform these models. Whether this is due to overfitting, where certain models capture dataset-specific cues, or simply because some models are better suited to particular datasets, this observation underscores the value of our multi-model approach. Its key strength lies in mitigating the limitations or over-specialization that any single model might exhibit.

\begin{figure*}[t!]
    \centering
    \makebox[\textwidth][c]{\includegraphics[width=13.5cm,keepaspectratio]{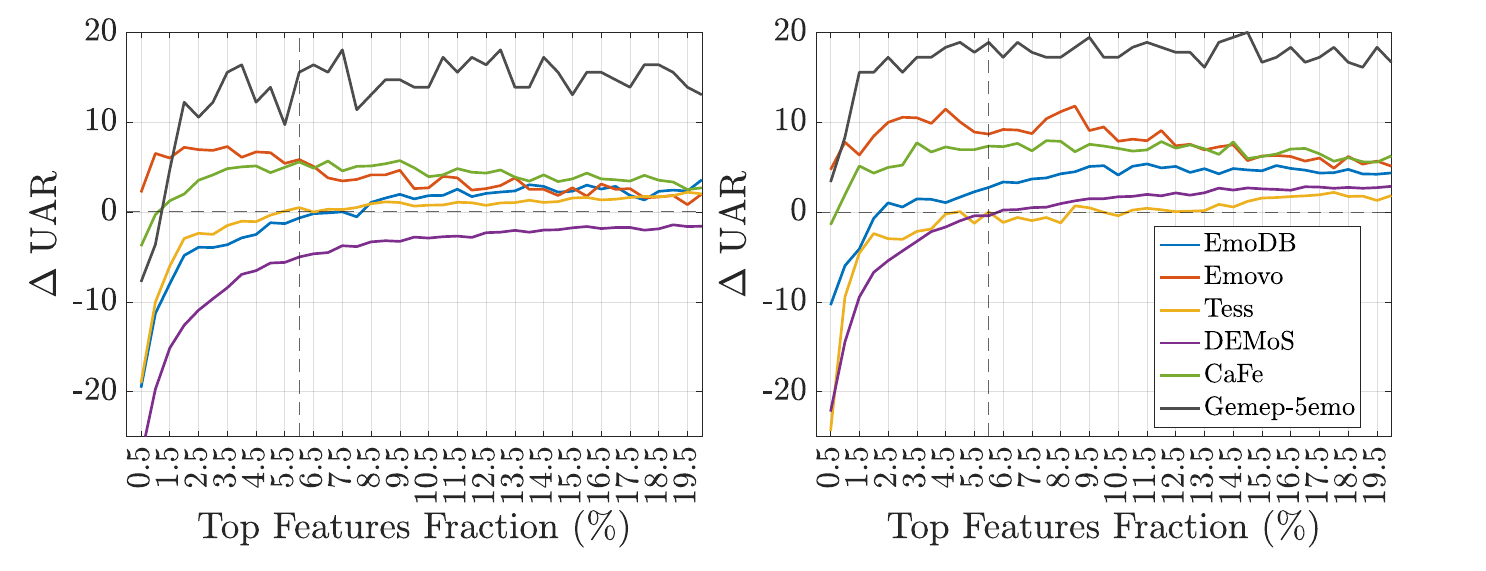}}
\caption{Average UAR difference across all models for each dataset when retraining models using top-ranked features at various thresholds, compared to baseline performance (i.e., when models are trained using all features). Left: Top-ranked features selected based on each model's own feature importance scores. Right: Top-ranked features selected using the proposed aggregated feature importance scores across all models.}
    \label{fig:fd2x3_uar}
\end{figure*}

\subsection{Feature selection: multiple vs one model}
\label{sec:fs}
To illustrate the advantage of our proposed approach for aggregating feature importance scores across models, in this section we compare the performance when models are trained under two conditions: (i) using top-ranked features derived from the aggregated importance scores across all models, as described in Section~\ref{sec:der}, and (ii) using top-ranked features based solely on each individual model’s importance scores.
These performances are compared to the baseline performance when all features are used.
As outlined in Section~\ref{sec: hyper}, each experiment is repeated three times.
Figure~\ref{fig:fd2x3_uar} shows the difference from the baseline performance for each dataset, averaged across all considered models\footnote{Note that the MLP and RBF-SVM models are also included in the average in the right figure. The outcome remains largely unchanged without MLP \& SVM. The marked performance enhancement isn't attributable to their inclusion.} using various thresholds of top-ranked features based on either the aggregated feature importance scores or the individual feature importance scores.
The presented results demonstrate the advantage of using aggregated feature importance scores across multiple models. Specifically, even with just the top-ranked 0.5\% features, performance is higher for each dataset when the features are selected based on aggregated importance scores, as compared to using individual model rankings. Furthermore, when selecting the top-ranked 6\% of features using aggregated importance scores, baseline performance is achieved for all datasets. In contrast, when selecting the top-ranked 6\% of features using individual importance scores for each model, baseline performance is achieved only for three datasets.

\subsection{Most important acoustic cues for SER}
\begin{figure}[t!]
\hspace{-1cm}
    \centering
\includegraphics[width=8.5cm,keepaspectratio]{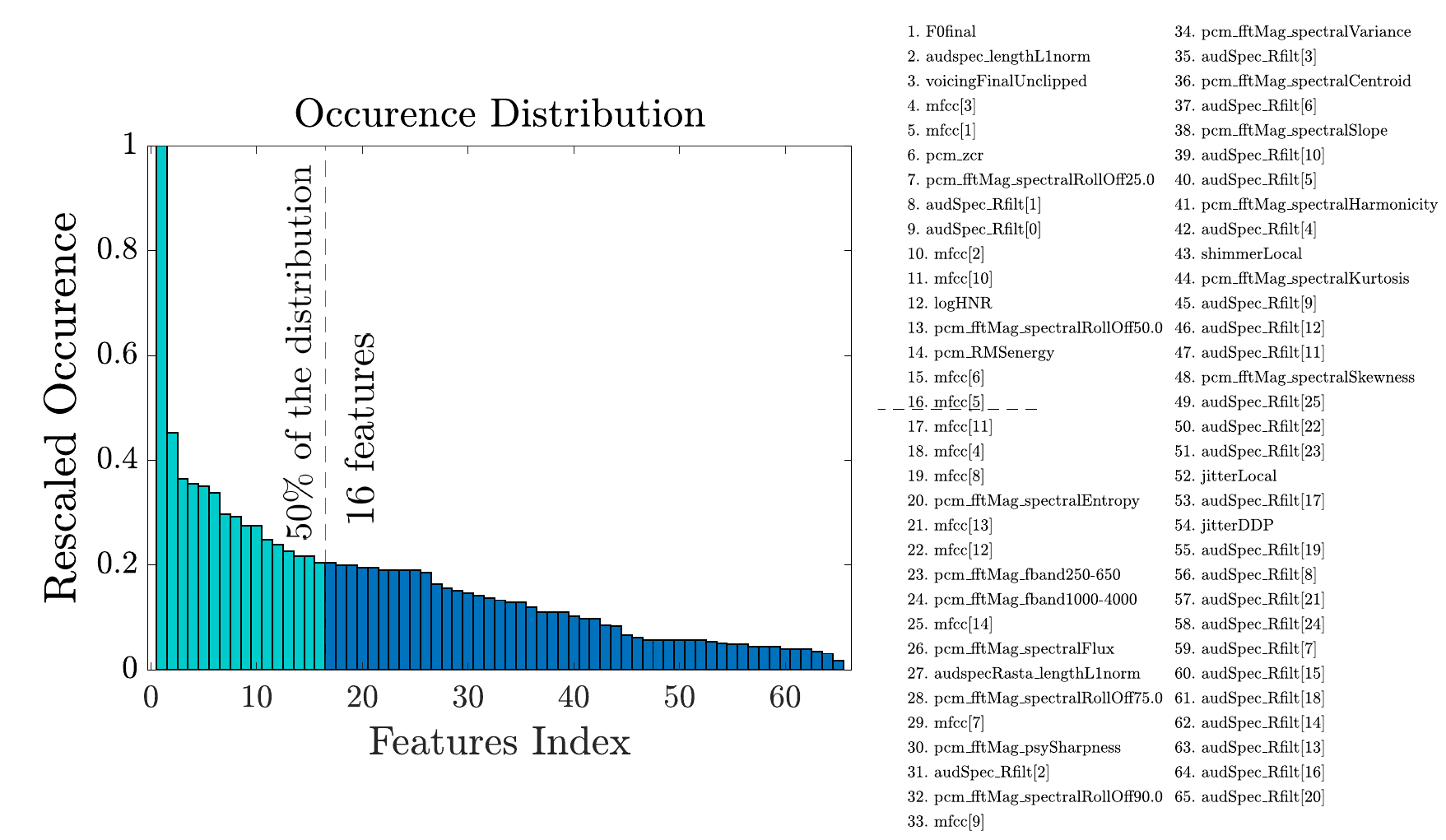}
\caption{Normalized occurrence of LLDs in the top-ranked $6$\% features.} 
    \label{fig:topllds}
\end{figure}

The results in Section~\ref{sec:fs} show that using $6$\% of top-ranked features through aggregated feature importance scores yields the same or better performance than using the complete feature set for all datasets (cf. Figure~\ref{fig:fd2x3_uar}).
Retaining these $6$\% of top-ranked features, we obtain a multi-set of $2,282$ features, among which $1,687$ are distinct.
This considerable number of distinct features underscores the complexity of the emotion recognition task, where information lies on a diverse and extensive set of features.
Since each feature stems from an LLD, we determine the importance of an LLD by counting the number of times features derived from it appear in the top-ranked features (with $\Delta$LLD features also attributed to the specific LLD it is extracted from).
Figure~\ref{fig:topllds} presents the normalized occurrence distribution of these LLDs. It can be observed that a small number of LLDs, such as \emph{F0final} and \emph{audspec\_lengthL1norm}, are highly important. 
While the entire list of LLDs is significant, as each LLD represents an important acoustic cue for SER, the distribution shows a sharp decline, followed by a gradual, steady slope. The cutoff point, indicated by the dashed line, represents 50\% of the total occurrences, corresponding to only $16$ LLDs. Although this cutoff is used for illustrative purposes and does not have intrinsic significance, it highlights the top $16$ LLDs that are particularly important. In the remainder of this section, we provide insights into these top $16$ LLDs, which are crucial for SER, as determined using our proposed approach.

\emph{F0Final}. \enspace The fundamental frequency, commonly referred to as pitch, stands out as the LLD on which models heavily rely for effective emotion recognition. This observation aligns with existing literature on emotion recognition (see Section~\ref{sec:acoustic}), where the significance of pitch is widely acknowledged.

\emph{AudSpec}. \enspace The auditory spectrogram is a transformation of the linear-frequency spectrogram that takes into account the non-linear frequency resolution of human hearing. 
Various characteristics of the auditory spectrogram as captured by the LLDs \emph{audspec\_lengthL1norm}, \emph{audSpec\_Rfilt[0]}, and \emph{audSpec\_Rfilt[1]}, appear to be important acoustic cues for SER. The LLD \emph{audspec\_lengthL1norm} is the magnitude of the $l_1$ norm of the auditory spectrum, broadly corresponding to the perceived loudness. The LLDs \emph{audSpec\_Rfilt[0]} and \emph{audSpec\_Rfilt[1]} are the first coefficients of the Rasta transformation used to make to make auditory spectrograms more resilient to noise, adverse conditions and other factors that can affect speech perception and analysis.

\emph{VoicingFinalUnclipped.} \enspace  This LLD represents the probability that F0Final is voiced. Emotional states can significantly affect phonation and voice intensity, both of which are key factors in vocal sound production. Since these factors directly influence whether F0Final is voiced, this variable is expected to vary with the speaker’s emotional involvement.

\emph{MFCCs.} \enspace Mel-Frequency Cepstral Coefficients (MFCCs) are derived from the cosine transform applied to lagirlogarithmic Mel scale representation of the input signal spectrum. These coefficients offer a compact representation of the overall spectral contour and are robust to variations in recording conditions and speaker characteristics.
The lower MFCC coefficients, which correspond to formants, tend to be more important for SER than the higher MFCC coefficients, which capture finer spectral details and rapid variations in the spectrum.

\emph{LogHNR.} \enspace The logarithm of the harmonic-to-noise ratio indicates voicing characteristics, and is hence, also relevant for SER.

\emph{PCM.} \enspace Apart from the aforementioned features, the last category of important features includes Pulse Code Modulation (PCM) features related to variation and energy derived from the signal or its spectrum. Even without incorporating human auditory specificity, these features are relevant for the SER task.
More specifically:
\begin{itemize}
    \item The \emph{pcm\_RMSenergy} measure is closely associated with loudness, which is an important acoustic cue for SER.
    
    \item Emotions have significant effects on spectral energy distribution. For instance, emotions like sadness are often associated with low harmonic energy above $1$ kHz, whereas emotions like anger, happiness, or fear, tend to manifest as high harmonic energy within this frequency range~\cite{guz13}. Hence, it's unsurprising to find several features linked to such spectral characteristics to be important for SER. These include the \emph{pcm\_fftmag\_spectralRollOff25.0}, which designates the frequency demarcating $25$\% of the signal's energy, and the \emph{pcm\_fftmag\_spectralRollOff50.0}, which designates the frequency demarcating $50$\% of the signal's energy. 
    \item  The (\emph{pcm\_zcr}) measure, which is the zero crossing rate quantifing the number of sign alternations in a signal, also ranks among the crucial features. Indeed, emotion-related changes in vocal expression, intensity, and timbre might affect the temporal variations captured by the zero crossing rate.
\end{itemize}

These are the characteristics emphasized by our study. The exact computation of these LLDs can be found in~\cite{eyben2015real}. The next phase of this project will involve developing a method to explicitly or implicitly preserve some of these features and determine the effectiveness of such an approach.

\section{Conclusion}\label{sec:xp}

In this paper, we have presented a comprehensive analysis employing multiple models and datasets to identify the most significant and robust features for the SER task. To the best of our knowledge, this is the first study to utilize such a diverse array of datasets and models to establish feature importance in the SER literature.
Despite the complexities inherent in SER, we have identified several key features that are highly relevant in this context. In the future, we will extend our findings to specific emotions by repeating the experiments using recall for each emotion as the evaluation metric. Additionally, we will investigate whether the degradation of these features under adverse conditions such as background noise, channel variations, or speaker diversity, correlates with the performance drop of automatic models and human perception of emotions.

\section{Acknowledgments}
This work was partially supported by Sonova AG, St\"{a}fa, Switzerland.

\bibliography{fa2025_template}

\end{document}